\documentclass[jcp,preprint,showemail,showpacs]{revtex4-1}
\usepackage{graphicx}
\usepackage{amsmath}
\usepackage{enumitem}
\usepackage{color}
\usepackage{soul}

\begin{document}

\title{A thermally-driven differential mutation approach for the structural optimization of large atomic systems}
\author{Katja Biswas}
\email{Katja.Biswas@ung.edu}
\affiliation{Department of Physics, University of North Georgia, Oakwood, GA 30566}

\date{\today}

\begin{abstract}
A computational method is presented which is capable to obtain low lying energy structures of topological amorphous systems. The method merges a differential mutation genetic algorithm with simulated annealing. This is done by incorporating a thermal selection criterion, which makes it possible to reliably obtain low lying minima with just a small population size and is suitable for multimodal structural optimization. The method is tested on the structural optimization of amorphous graphene from unbiased atomic starting configurations. With just a population size of six systems, energetically very low structures are obtained. While each of the structures represents a distinctly different arrangement of the atoms, their properties, such as energy, distribution of rings, radial distribution function, coordination number and distribution of bond angles are very similar.
\end{abstract}

\pacs{2.60.Pn, 2.70.-c, 61.48.Gh}

\maketitle

\section{Introduction}
Obtaining computational models of structures of amorphous graphene is currently gaining traction, as more and more experimental methods to generate amorphous graphene structures become available. Experimentally it has been shown that amorphous graphene can be obtained by prolonged Ga-ion beam irradiation\,\cite{Zhou2010,KotakoskiNano2015}, prolonged electron-beam irradiation\,\cite{Teweldebrhan2009,KotakoskiPRL2011}, chemical vapor deposition\,\cite{Zhao2012, Li2013} and unzipping of amorphous carbon nanotubes\,\cite{Chattopadhyay2014}. 

I will introduce a method which is inspired by the differential mutation (DM) algorithm, a special case of the differential evolution (DE) algorithm introduced by Price and Storn\,\cite{Storn1995,Stornbook2005}. A recent review about the DE algorithm can be found in \cite{Das2016}. The DE-algorithm is a genetic algorithm that works on the principle of mutation, cross-over and selection.
To work satisfactorily, the original DE method requires a large population size in order to have sufficient variation in the possible solution space\,\cite{Stornbook2005}. This poses a problem for structural optimization of large atomic systems due to the following reason. A population consists of multiple versions of the atomic system, each representing a different structural arrangement of the atoms. 
If the configuration in 3-dimensional space is the focus of the optimization, this would require that $3\times N_p \times N$ variables need to be optimized, where $N_p$ is the number of atoms and $N$ the number of systems.
Since $N$ has to be large enough to include the possible solution space, one can easily see that such an approach might be beyond current computational limits for large systems. 

To overcome this problem, I merged the DM algorithm with a thermal selection criterion inspired by simulated annealing (SA). This allows for thermally bounded increases in energy, which can lead to the breaking of bonds. Particularly during the early stages of the optimization, this allows for rearrangements of the atoms corresponding to the escape from unfavorable local minima traps. 
Further, to speed up the computation, the cross-over (part of the original DE method) has been omitted as it would lead to unnecessary complications in the optimization procedure originating from the initial random placement of the atoms within the individual systems. In fact, including cross-over would require the atoms to be sorted based on their relative location in space, which then would have to be frequently updated, in order to avoid the creation of energetically unfavorable holes and overlap structures.

Developing hybrid algorithms incorporating DE and SA is relatively new and has so far only be implemented and tested on benchmark functions and circuit design.
For circuit optimization problems, Olen{\v{s}}ek et. al developed a parallel simulated annealing and differential evolution (PSADE) algorithm\,\cite{olensek2011}, which was later modified by combining the algorithm with a population based ranking (DESAPR)\,\cite{olensek2016}. Combining the DE algorithm with SA was also used in the development of  
 ESADE (enhanced self-adaptive differential evolution)\,\cite{Guo2014} and iSADE\,\cite{Zhao2013}, which combines a hybrid DE with SA and self-adaptive Gaussian immune operation. Tests on benchmark functions showed a very good performance and robustness of the hybrid algorithms. 
However, implementation and augmentations of these methods to large scale structural optimization problems have not yet been done. Optimization problems which deal with the structural arrangement of atoms are different than the range of problems tackled with the above mentioned procedures.\\
Zacharias\,\cite{Zacharias1998} developed an algorithm that switches between a pristine genetic algorithm and a pristine simulated annealing procedure and applied it to small silicon clusters.

Applications of genetic algorithms to nanocluster and crystal structure optimization has shown tremendous successes\cite{WoodleyReview}.
Particularly in the field of crystal structure prediction Woodley and Catlow developed a genetic algorithm that only uses the knowledge of the dimensions of the unit cell\,\cite{Woodley1999, Woodley2004_1,Woodley2004_2,Woodley200984}.
Oganov used his USPEX code (USPEX: Universal Structure Predictor: Evolutionary Xtallography), which merges an evolutionary algorithm with ab-initio calculations, to predict crystal structures using supercells\,\cite{Oganov2006,GLASS2006713,Oganov2008}.
Both the methods make use of the repetitive nature of atomic configurations in crystals, as such they can achieve high quality minimum structures  by using only a very limited number of atoms which are then optimizated representing unit or supercells.
However, different from crystalline materials, amorphous materials to not exhibit a long range order. Hence assumptions about repeated spatial arrangements of the atoms cannot be used and the number of atoms making up the size of the system has to be large enough in order to avoid finite size effects on the properties of the final structure.

The goal of this paper is to make the DE/DM fit for structural optimization problems of amorphous systems, which require a large number of atoms, 
are high dimensional  and for which the minima are degenerate originating from the indistinguishability of the atoms. 
The proposed method resembles a thermally-driven differential mutation. Using an analytic expression for the potential energy stored in the configurations of the atoms and a confining volume, it is capable to reliably obtain low lying energy minima with just a very limited population size and can be used for multimodal structural optimization problems involving large 
amorphous systems in real-space representation.
As will be illustrated on the optimization of amorphous graphene, the advantage  of the method is that it provides with multiple distinctly different configurations which are close in energy and have very similar features in the radial distribution function, the occurrence of ring sizes, the bond angle distribution and the coordination number.

The method provides a new approach for the computational discovery of amorphous graphene structures, which can be used in further computational studies. Up-to-date only very few computational methods have been used to this effect. Amorphous graphene structures were obtained by creating Stone-Wales defects\,\cite{Stone1986} using the Wooten-Wearie-Winer\,(WWW) method\,\cite{WWWmethod}. 
Starting from pristine graphene structures, Kapko\,\cite{Kapko2010}, Tuan\,\cite{Tuan2012} and Kumar\,\cite{Thorpe2012} used this method to generate two-dimensional structures. Whereas Li and Drabold\,\cite{Li2011,LiandDrabold2013}, and Mortazavi\,\cite{Mortazavi2016} obtained three-dimensional structures. Small three-dimensional amorphous graphene layers have also been obtained by Popescu\,\cite{Popescu2013}.
Structures were also obtained by cooling from high temperatures using molecular dynamics (MD) by Van Hoang\,\cite{VanHoang2015} and Kumar\,\cite{Thorpe2012} and 
Holmstr\"om\,\cite{holmstrom2011} used a stochastic quenching method. Relaxation via MD and time-stamped force-bias Monte-Carlo methods were used too to observe healing in disordered amorphous graphene structures\,\cite{Bal2014}. 

The article is structured in the following way. In section 2 the method is introduced. In section 3, the algorithm is then applied to generate amorphous graphene structures for systems consisting of 500 atoms and a population size of only six systems. Computational details will be given in this section and the results are compared with those from refs.\,\cite{Tuan2012,Thorpe2012,Li2011,LiandDrabold2013,Mortazavi2016,Popescu2013,VanHoang2015,holmstrom2011}.

\section{Method}

\begin{figure*}[t!]
\begin{minipage}{0.25\textwidth}
\includegraphics[width=1.75in]{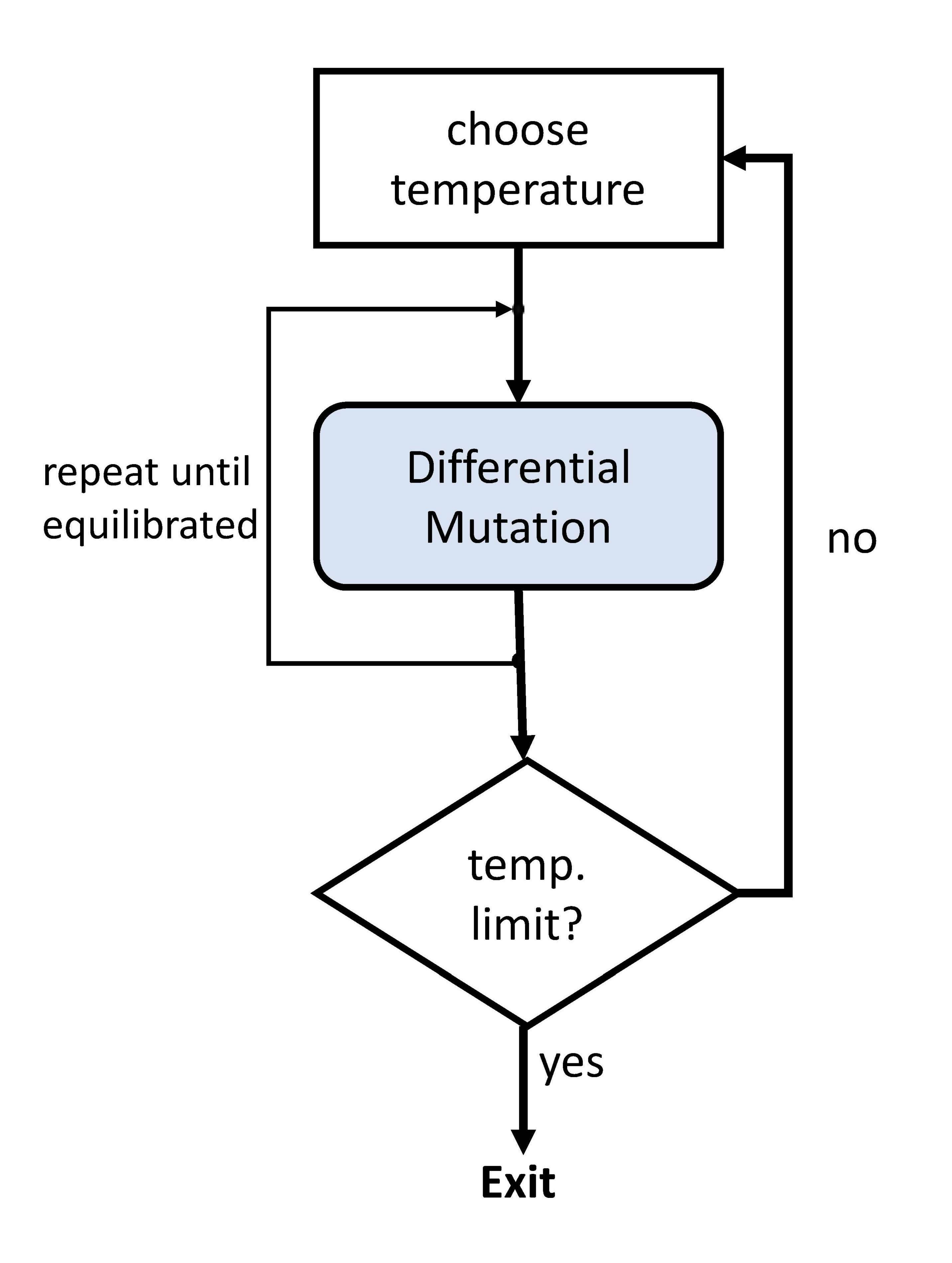}
\caption{\label{fig1} Schematic drawing of the algorithm. The modified differential mutation algorithm is shown in fig.\,(\ref{fig2}) }
\end{minipage}
\hfill
\begin{minipage}{0.725\textwidth}
\includegraphics[width=5.5in]{figure2}
\caption{\label{fig2} Schematic drawing of the modified differential mutation algorithm with embedded thermal selection criteria.}
\end{minipage}
\end{figure*}

Figs.\,(\ref{fig1}) and (\ref{fig2}) show the schematics of the algorithm. At the start of the evolution sequence a set of $N$ systems $\{\zeta_1,\zeta_2,\ldots ,\zeta_N\}$  are initiated. For each system $N_p$ atoms are randomly distributed confined to some specified volume. This has to be done in such a way that the starting configuration of the atoms is different for each system. An initial temperature $T$ and the differential mutation parameters, $C_r$ and $F$, are chosen. Ideally, if known, the initial temperature should be higher than the melting point of the corresponding crystalline system to ensure sufficient mobility at the beginning of the optimization run. $C_r\in[0,1]$ influences the number of site mutations of the individual systems and $F$ determines the weight of the difference vector (see below) in mutations. $F$ and $C_r$ should be chosen such as to achieve a good acceptance/rejection ratio of the mutant trial vectors. \\
After the initialization, the differential mutation routine is started. From the $N$ systems three systems are selected of which one is the target system $\zeta_l$ and the other two $\zeta_m$ and $\zeta_n$ will be used to mutate the target system via weighted difference. This is done in the following way, from each of the two systems $\zeta_m$ and $\zeta_n$ one atom (say $r_i \in \zeta_m$ and $r_j \in \zeta_n$) is chosen randomly. 
The difference in location of the atoms (i.e. $\Delta r= r_i-r_j$) is calculated and added, with the difference weight factor $F$, to the position of an atom $r_k$ from the target system $\zeta_l$, i.e. $\tilde{r}_k=r_k+F*\Delta r$. This creates a mutation on the position of atom $r_k$. The procedure is repeated following the rule, that at least the position of one atom, say $r_{\tilde{k}}$, will change. This atom is determined via random selection from all the atoms of $\zeta_k$. For the other $N_p-1$ atoms a probability criterion is used where the parameter $C_r$ determines the probability of mutation. For this purpose, for every atom ($r_k$ with $k \neq \tilde{k}$), a random number is drawn from a uniform distribution in the range zero to one. If this number is less than $C_r$ the location of the atom is mutated according to the mentioned procedure otherwise the original location of this atom is carried into the mutant system.
\begin{figure*}[t!]
\begin{minipage}{0.475\textwidth}
\includegraphics[width=3.25in]{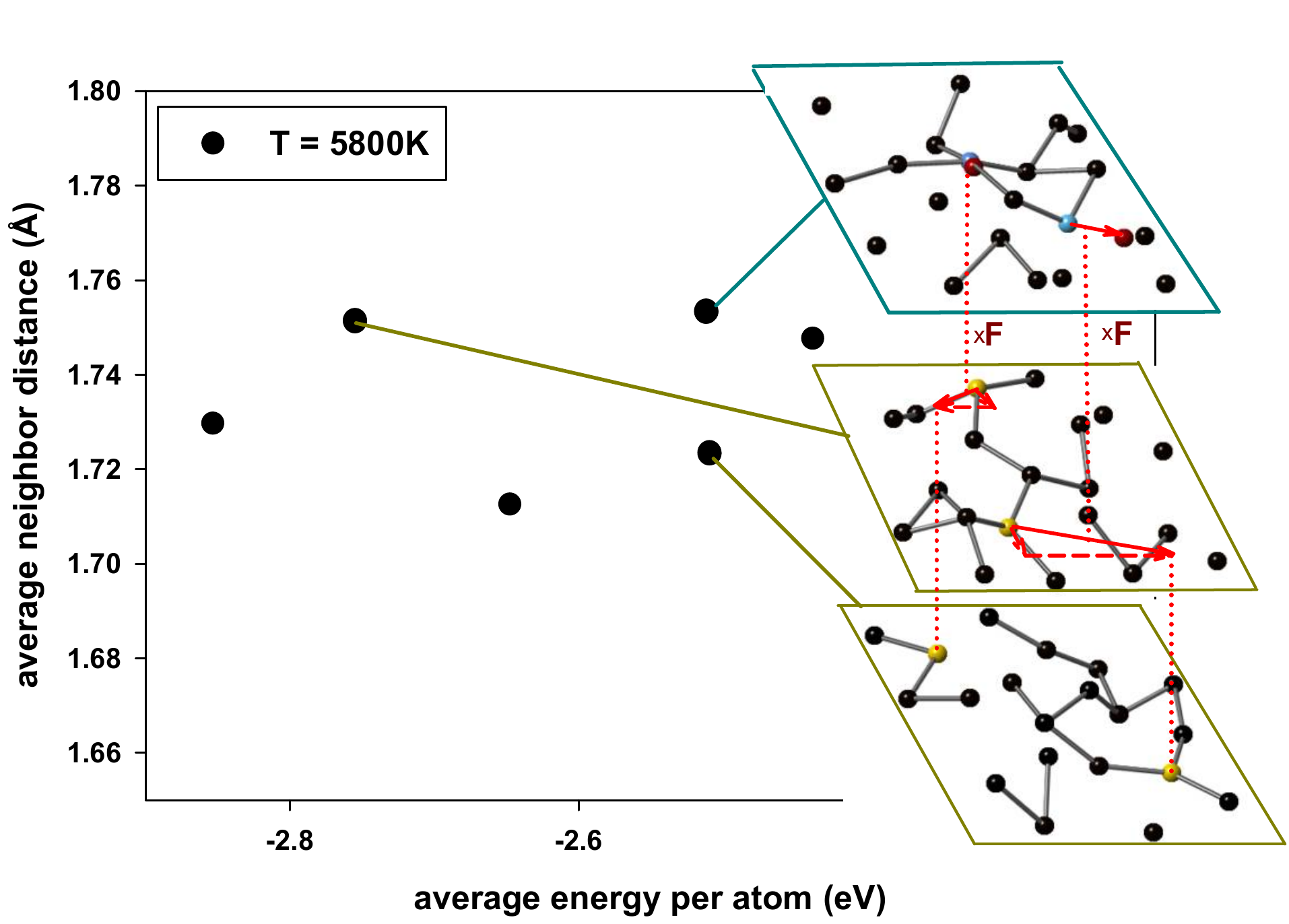}
\caption{\label{fig3} Illustration of the algorithm at a population of 6 systems (black dots) each consisting of 20 carbon atoms. The magnifications illustrate the target system (outlined in blue) and the two systems used in determining the difference (outlined in yellow). The difference is added to the target system to create mutations on the locations of some of its atoms. }
\end{minipage}
\hfill
\begin{minipage}{0.475\textwidth}
\includegraphics[width=3.25in]{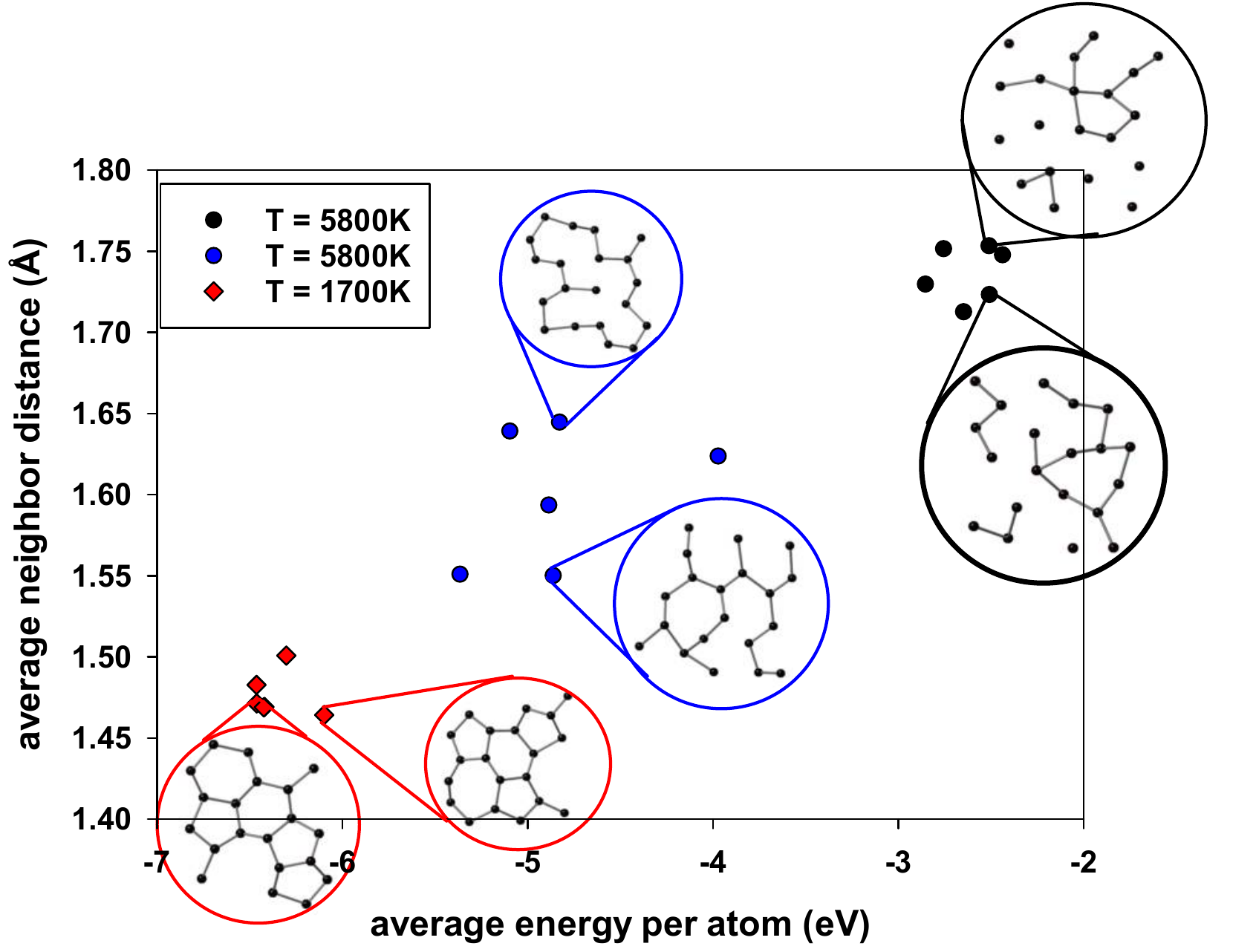}
\caption{\label{fig4} Illustration of the evolution of the population of 6 system consisting of 20 carbon atoms each. Illustrated are the systems after initialization at $T=5800K$ (black diamonds), after equilibration at the same temperature (blue) and at a lower temperature (red). The magnifications show examples of the individual systems.}
\end{minipage}
\end{figure*}
Once the mutant system $\tilde{\zeta}$ has been obtained its potential energy $\tilde{U}$ is compared to that of the target system $U$. If the difference $\Delta U=\tilde{U}-U$ is smaller or equal than zero, the mutant system is accepted as member of the next generation otherwise the thermal metropolis criterion is used. For this purpose a random number from a uniform random number distribution in the range zero to one (i.e. $rng[0,1]$) is drawn and compared to the Boltzmann factor of the change in energy. If  $rng[0,1]<\exp(-\beta \Delta U)$ the mutant is accepted otherwise not, in which case the original target system is carried into the next generation. Here $\beta =1/(k_BT)$ and $k_B$ is the Boltzmann constant. This process is repeated for the remaining $N-1$ systems $\{\zeta_i\}_{i\ne k}$, specifying each of the remaining systems as target system and randomly choosing two more systems from the old generation as difference systems. The resulting population is the new population.\\
The procedure is repeated until equilibration at the specified temperature is reached at which point the temperature is lowered and the process repeated. 
Equilibration is reached once the average of the energy of the individual systems does not change any longer significantly. The individual values in energy will still fluctuate within some bounds, indicative of the transitions of the systems between different energy basins. In general the upper bound of the energy fluctuations are higher at higher temperatures than at lower. The optimization routine can be stopped when either sufficient convergence in the location of the atoms of each of the systems is reached or at zero temperature. Here, sufficient convergence means that the location and energies of the single atoms in the systems $\{\zeta_1,\ldots ,\zeta_N\}$ do not change any longer significantly, i.e. the system reached the vicinity of a low energy minimum from which it can no longer escape. In this case other methods such as gradient optimization or zero temperature Monte-Carlo optimization may be used to free the systems of the temperature induced vibrational disorder.

The procedure can be summarized as follows:
\begin{enumerate}[label=(\roman*)~~~~]
\item\label{step1}{$\bullet$ choose parameters $F$, $C_r$ and starting temperature $T$}
\item\label{step2}{$\bullet$ initialize the population systems $\{\zeta_1,\zeta_2,\ldots ,\zeta_N\}$}
\item\label{step3}{{FOR $\ell =1$ to $N$ ~~~ !\# $\ell$ is the target system}
 \begin{enumerate}[leftmargin=0.4cm,labelwidth=1.2cm,start=4,label=(\roman*)~~~~~~~~]
\item\label{step4}{$\bullet$ randomly select systems $\zeta_n$ and $\zeta_m$ (with $n\neq m$, $m\neq \ell$ and $n\neq \ell$)}
\item\label{step5}{$\bullet$ draw a random number $k_0$ from $\{1,2,\ldots ,N_p\}$}
\item\label{step6}{{FOR $k=1$ to $N_p$   ~~~ !\# $k$ is an atom of $\zeta_\ell$}}
\item\label{step7}{{~~~IF $k=k_0$ or $rng[0,1]\le C_r$}
  \begin{enumerate}[start=8,label=(\roman*)~~~~~~~~~~~~~]
\item\label{step8}{$\bullet$ randomly select atom $r_i$ from system $\zeta_m$ and atom $r_j$ from system $\zeta_n$}
\item\label{step9}{$\bullet$ form the difference vector $\Delta r=r_i-r_j$}
\item\label{step10}{$\bullet$ mutate the location of atom $k$ using the weighted difference vector $\Delta r$, i.e. $\tilde{r}_k=r_k+F*\Delta r$}
  \end{enumerate}
{~~~ELSE}
 \begin{enumerate}[start=11,label=(\roman*)~~~~~~~~~~~]
\item{$\bullet$ set $\tilde{r}_k=r_k$}
\end{enumerate}
{~~~ENDIF\\
ENDFOR}}
\end{enumerate}
\begin{enumerate}[start=12,label=(\roman*)~~~~~~~~]
\item\label{step12}{$\bullet$ calculate the difference in energy between the mutant and the target system, i.e. $\Delta U=\tilde{U}-U$}
\item\label{step13}{$\bullet$ if $\Delta U\le 0$ or $rng[0,1]< exp[-\beta\Delta U]$ accept the mutant into the next generation, otherwise carry system $\zeta_l$}
\end{enumerate}
{ENDFOR}}
\end{enumerate}
\begin{enumerate}[start=14,label=(\roman*)~~]
\item\label{step15}{$\bullet$ check for equilibration; if equilibration at temperature $T$ is reached continue at next step, otherwise go to 
\ref{step3}}
\item\label{step16}{$\bullet$ if $T\le T_o$ or some other convergence criteria is reached exit the program, otherwise lower the temperature and go to \ref{step3}}
\end{enumerate}

Figs.\,(\ref{fig3}) and (\ref{fig4}) illustrate the procedure at the structural optimization of amorphous graphene consisting of 20 atoms using a population size of $N=6$ systems. The systems were initialized by placing atoms at random positions in a confined volume [see insets of Fig.\,(\ref{fig3})]. The starting temperature was chosen as $T=5800K$. In order to illustrate the difference between the systems, fig.\,(\ref{fig3}) shows a plot of the average distance between nearest neighbors versus the average energy per atom at $T=5800K$. The insets illustrate step (viii) to step (x) of the procedure. From each of the two difference systems (yellow insets)  two atoms were selected at random and the difference in their positions was added using the weighting factor $F$ to the atoms of the target system (blue inset). This created mutations (red atoms) on the positions of two of the target atoms (blue atoms).\\ 
Fig.\,(\ref{fig4}) shows plots of the average distance between nearest neighbors versus the average energy per atom for the systems at different stages during the evolution process. The black dots represent the systems directly after their initialization. The blue dots and blue insets show the systems after the initial equilibration, and the red dots represent the configurations of the systems at a much lower temperature. As can be seen, at high temperatures after equilibration the systems sample a wider range of energies than at low temperatures. This is to be expected since the available configuration space is larger at higher temperatures than at lower, allowing for more variety in the mutation. Whereas at low temperatures the systems become more and more trapped in structural arrangement that are similar in energy, but yet represent different configurations.

\section{Results}
\begin{figure}[h!]
\includegraphics[width=1.4in]{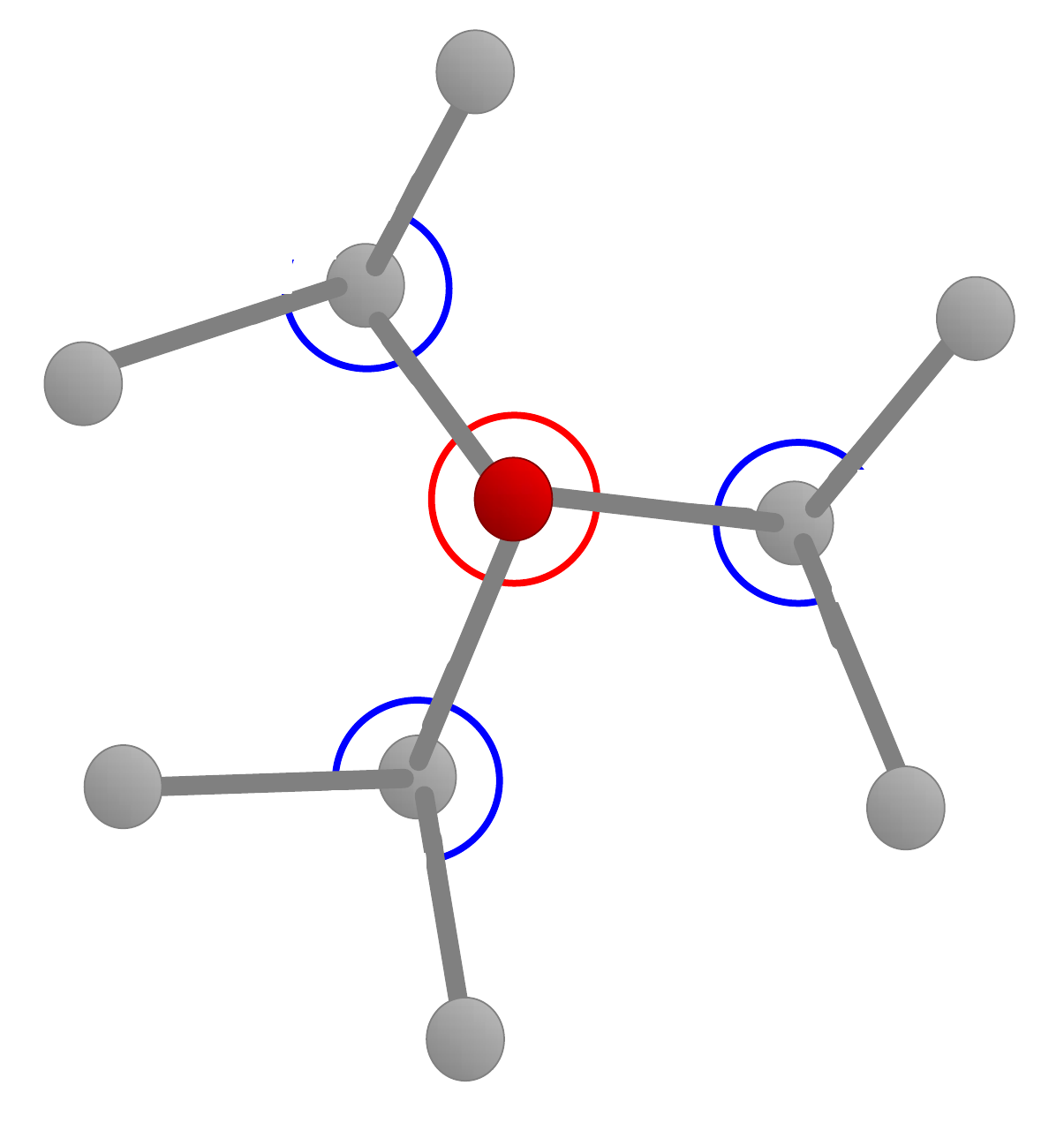}
\caption{\label{nnapproxfig} To speed up the calculation only the angles with atom $i$ (red atom) at the vertex (drawn in red) were used for the calculation of the three-body contributions in the energy.}  
\end{figure}

Low energy structures of amorphous graphene were obtained using the bond-order potential introduced by Erhart and Albe\,\cite{Erhart2005}. 
To speed up the computation the thermally driven differential mutation procedure was performed using a center approximation in the calculation of the effective three-body energies.
This approximation is depicted in fig.\,(\ref{nnapproxfig}). For each displacement of an atom, say atom $i$ (depicted in red), the bond angle potential was calculated taking into account only the angles with atom $i$ at their vertex (shown in red), but not the angles for which atom $i$ defines one of the sides (shown in blue). The pairwise distance dependent interactions were not modified. Since the calculation of three-body interactions is computationally the most costly, this small approximation significantly reduced the computation time. The gradient procedure was performed using the full Erhart and Albe\,\cite{Erhart2005} potential. 

Six systems $\{\zeta_1 ,\ldots ,\zeta_6\}$ were initiated by placing for each of the systems 500 atoms at random positions within the confined volume of a box of dimensions $36.274 \text{\AA}\times 36.274\text{\AA} \times 1.400\text{\AA}$, corresponding to an atom surface density of $0.38 \text{\AA}^{-2}$. In x- and y-direction periodic boundary and minimum image conditions were imposed. In z-direction only periodic boundary conditions were implemented, which forces the atoms to arrange in a single layer sheet. The starting temperature of the system was chosen as $T=5800K$. This temperature value is well above the melting temperature of graphene, allowing easily for structural breaking and reformation of bonds. $C_r$ was set to $0.02$, giving an average of $11$ side mutations per generation for each of the systems.  $F$ was kept variable in the range $0.001$ to $0.005$. The values for $C_r$ and $F$ were determined from short trial runs in an effort to achieve a balance between sufficient side mutations per generation, a good acceptance rate of the mutant systems and the capability to escape local traps during the initial stages of the optimization.
The temperature was lowered following an  iterative exponential decay procedure $T_n=T_{n-1}\exp [-\gamma (n-1)]$, where $\gamma=0.01$ determines the rate of cooling and $n$ enumerates the cooling steps. After every change in temperature the systems were given enough trials to equilibrate, approximately $100,000$ trial displacements per atom.\\

The thermally driven DM optimization was performed until the systems were cooled to a temperature of $T= 75K$.
The obtained structures were then refined and relaxed to zero temperature using a standard gradient method\cite{Gradient2006}. For this purpose the force acting on the atoms was used to calculate their new positions according to $\vec{r}_i^k=\vec{r}_i^{k-1}+\alpha \vec{F}_i$. Where $\vec{r}_i^0$ are the positions of the atoms after the DM hybrid optimization, $\vec{F}_i$ denotes the
 net force acting on atom $i$ and $\alpha$ is a factor determining the step length of the procedure. It was found that a parameter of $\alpha=0.001$ led to a sufficiently fast convergence without enabling the systems to escape their current configurations.
The gradient optimization was performed without restrictions on the z-direction, allowing the surface to freely buckle. Further, the full C-C potential\,\cite{Erhart2005} was used (i.e. without the approximation in the three-body contribution to the potential). The gradient optimization was stopped once the energy of the systems converged.

\begin{figure}[b!]
\includegraphics[width=3.25in]{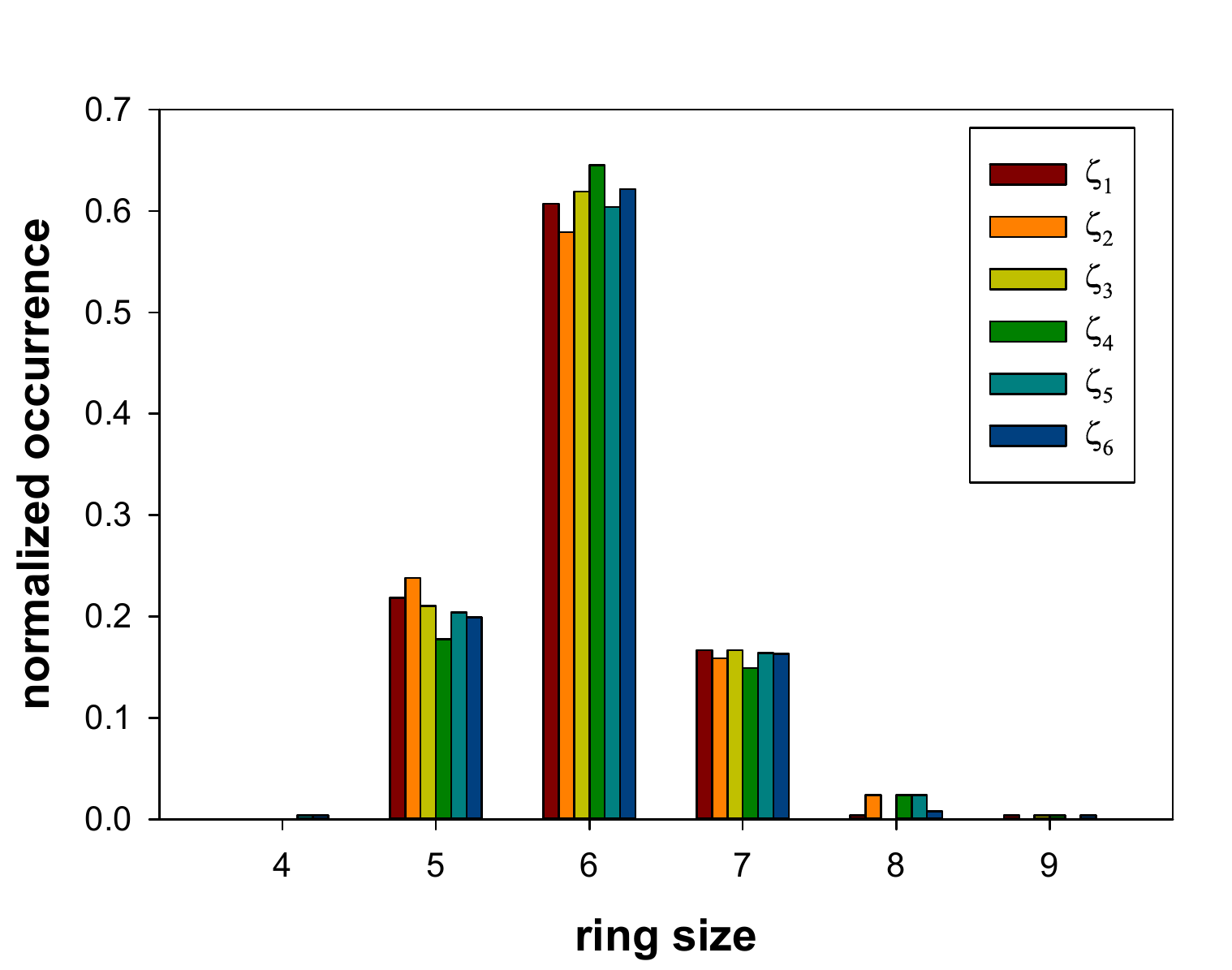}
\caption{\label{rings} Occurence of n-ring structures for the different systems.}
\end{figure}

\begin{table}
\caption{Root mean square deviation of the bond angle $\theta_{rms}$ from the mean and second moment $\mu_2$ of the ring distribution of the six systems $\zeta_1,\ldots ,\zeta_6$ compared to those from refs.\cite{Kapko2010,Tuan2012,Thorpe2012,VanHoang2015}.} 
\label{table2}
\begin{tabular}{|c|c|c|}
\hline
~ & $\theta_{rms}$ ($^o$) &  $\mu_2$\\ \hline\hline
$\zeta_1$ & $5.50$ & $0.44$ \\
$\zeta_2$ & $5.80$  & $0.49$\\
$\zeta_3$ & $5.26$  & $0.41$ \\
$\zeta_4$ & $5.37$  & $0.46$\\
$\zeta_5$ & $5.58$ & $0.48$  \\
$\zeta_6$ & $5.67$   & $0.45$\\ \hline
Kapko\cite{Kapko2010} (WWW) & $16$  & $0.4$  \\
Tuan\cite{Tuan2012} (WWW) &
\begin{tabular}{@{}c@{}} $11.02$ \\  $18.09$ \end{tabular} & 
\begin{tabular}{@{}c@{}} $0.47$\\ $0.88$ \end{tabular}\\
Kumar\cite{Thorpe2012}
\begin{tabular}{@{}c@{}} (WWW)\\   (MD) \end{tabular} & 
\begin{tabular}{@{}c@{}} ~$9.62-11.95$~~\\  $9.71-12.83$ \end{tabular} & 
\begin{tabular}{@{}c@{}} ~$ 0.43-0.67$~~\\ $0.45-0.65$\end{tabular} \\
 Van Hoang \cite{VanHoang2015} (MD) &  
\begin{tabular}{@{}c@{}}  $*$\\  $*$ \end{tabular} & 
\begin{tabular}{@{}c@{}} $2.475$\\ $1.919$\end{tabular}\\
 \hline
\end{tabular}
\end{table}

\begin{figure*}[ht!]
\includegraphics[width=4.75in]{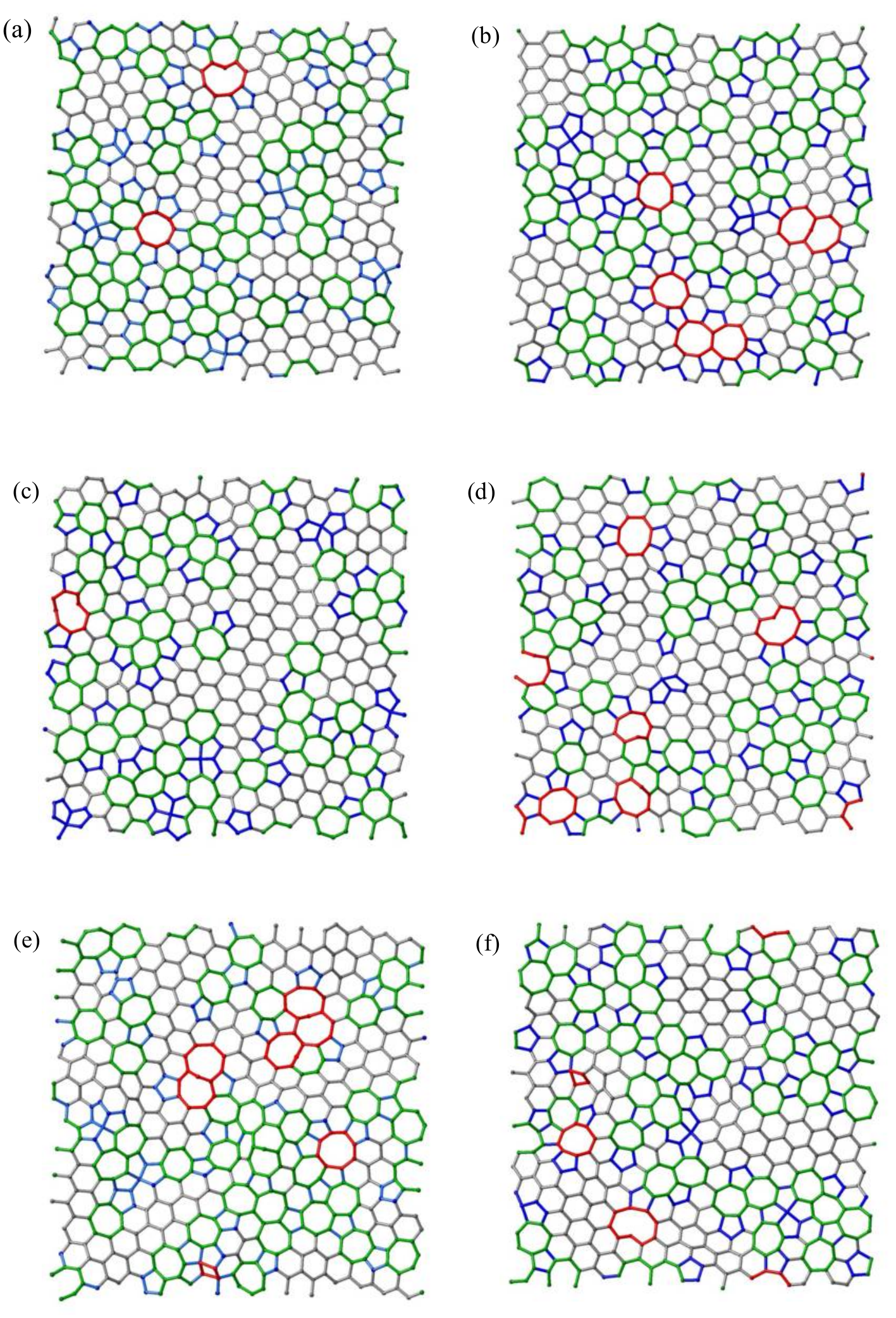}
\caption{\label{topview} {Topview of the systems $\zeta_1, \ldots ,\zeta_6$ labeled (a) through (f) respectively.
The bonds of the rings are colored based on the number of
 ring members, i.e. 5-ring (blue), 6-ring (gray), 7-ring (green), and  8-, 9- and 4-ring (red).}}
\end{figure*}

\begin{figure*}[ht!]
\begin{minipage}{0.45\textwidth}
\includegraphics[width=3.25in]{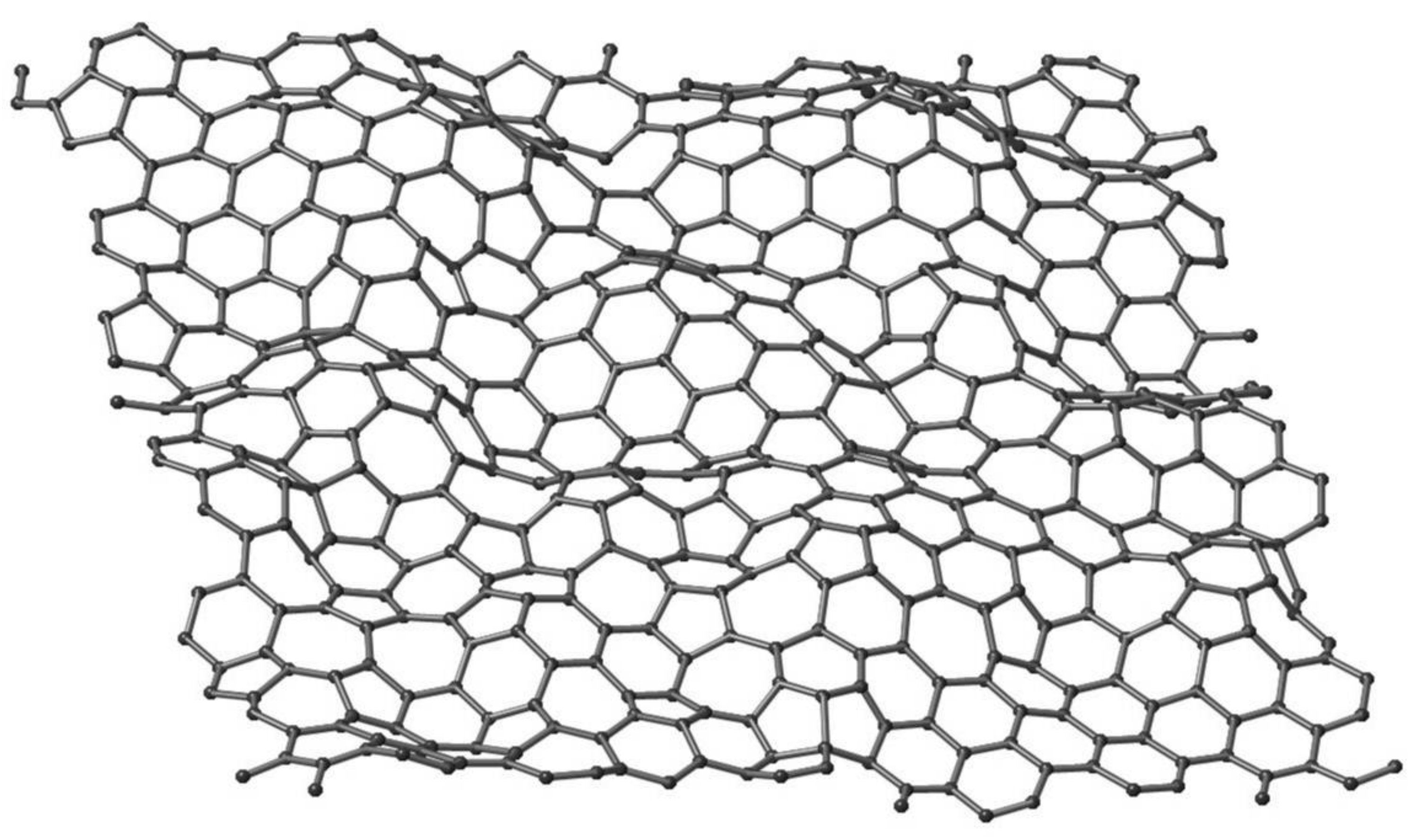}
\end{minipage}
\hfill
\begin{minipage}{0.45\textwidth}
\includegraphics[width=3.25in]{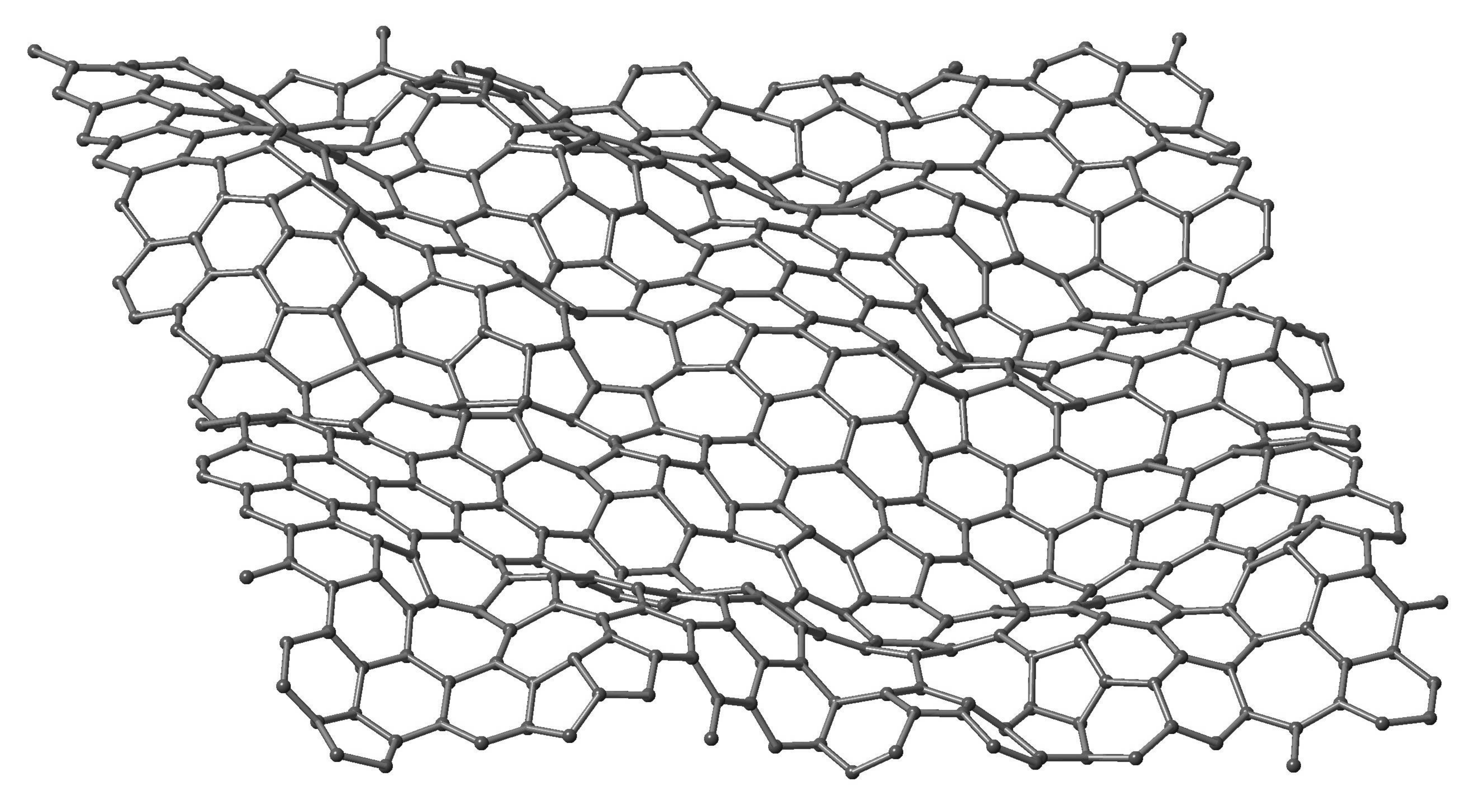}
\end{minipage}
\caption{\label{perspective} Perspective view of the structures $\zeta_1$ and $\zeta_2$.}
\end{figure*}

\begin{figure}[h!]
\includegraphics[width=3.25in]{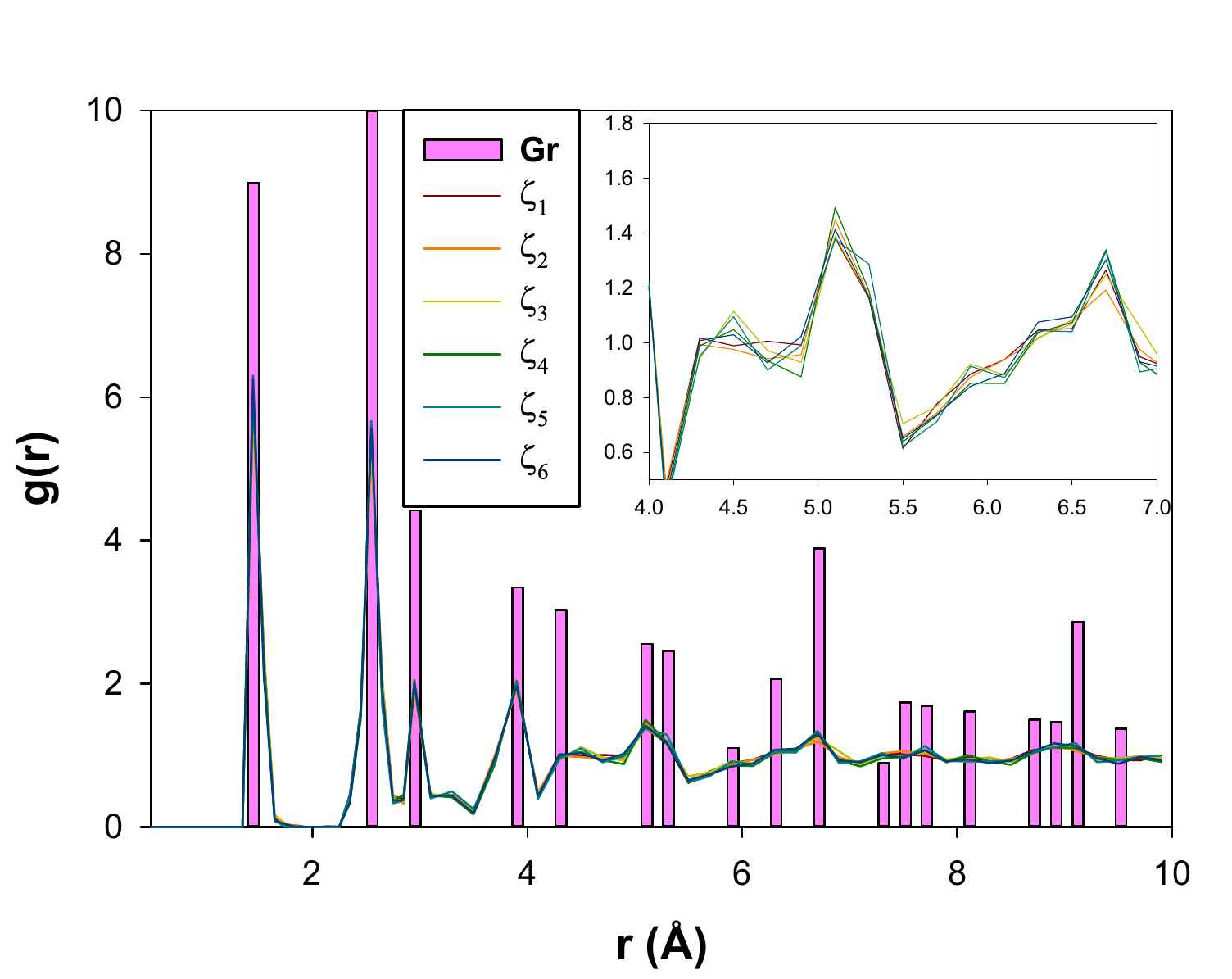}
\caption{\label{radial} The radial distribution function of the different systems from 0 to 10 \AA. The inset shows a magnification of the range $4.5$ to $8$ \AA.}
\end{figure}

\begin{figure}[h!]
\includegraphics[width=3.25in]{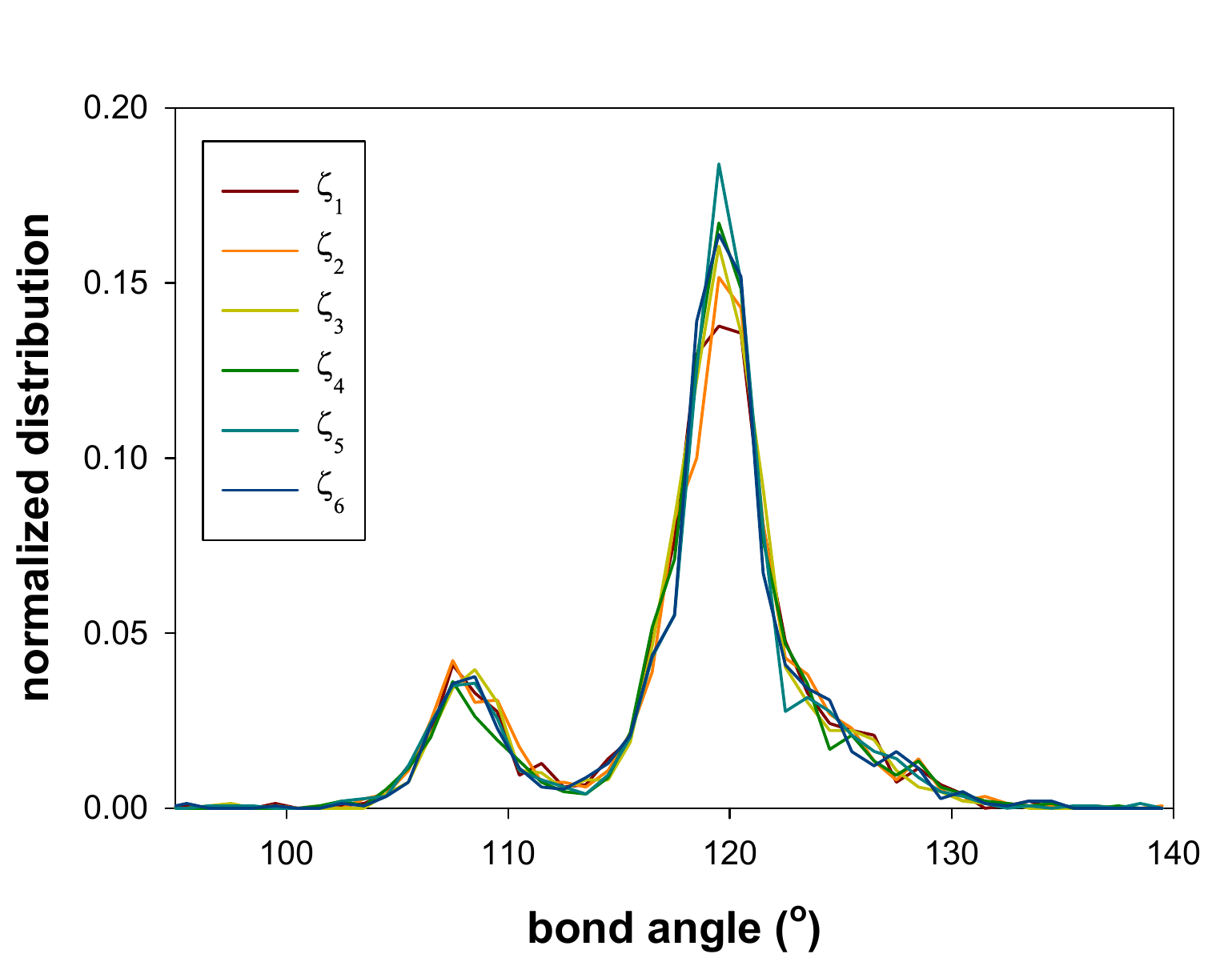}
\caption{\label{bondangle} The distribution of bond angles of the different systems.}
\end{figure}

The results can be seen in figs.\,(\ref{rings}) to (\ref{bondangle}) and tab.\,(\ref{table2}-\ref{table1}). Fig.\,(\ref{topview}) shows a top view of the final structures of the systems $\zeta_1,\ldots,\zeta_6$, and fig.\,(\ref{perspective}) shows, as example, the perspective view of the systems $\zeta_1$ and $\zeta_2$. It can be clearly seen that these structures represent different configurations. 

 The ring distribution can be seen in fig.\,(\ref{rings}), where the occurrence of the rings is plotted versus the ring size for the six systems $\{\zeta_1,\ldots ,\zeta_6\}$.
{ Similar to the structures obtained in refs.\,\cite{Thorpe2012,Tuan2012,Kapko2010,Li2011} 5-, 6- and 7-atom rings were dominating.}
 In all the systems about $58-65\%$ of the rings were 6-atom rings followed by 5-atom ($18-24\%$) and 7-atom ($15-17\%$) rings.
 Systems $\zeta_2$, $\zeta_4$ and $\zeta_5$ had each six 8-atom rings, giving a relative occurrence of $2\%$. The occurrence of 8-atom rings for the systems $\zeta_1$ and $\zeta_6$ was $0.4\%$ and $0.8\%$ respectively. System $\zeta_3$ did not exhibit an 8-atom ring. Only systems $\zeta_1$, $\zeta_3$, $\zeta_4$ and $\zeta_6$ had each one 9-atom ring, at $0.4\%$ occurrence.
{  Structures with up to 9-membered rings were also observed in studies by Holstr{\"o}m et al.\,\cite{holmstrom2011} and Van Hoang\,\cite{VanHoang2015}.}
 A 4-atom ring could only be found for systems $\zeta_5$ and $\zeta_6$, giving an occurance of $0.4\%$. { 
 In studies\,\cite{Thorpe2012,Tuan2012,Kapko2010,Li2011,holmstrom2011} there were no rings reported with less than 5-members, and only Van Hoang\,\cite{VanHoang2015} reported structures of 3- and 4-membered rings with $0.020\%-0.039\%$ and $0.237\%-0.393\%$ occurance respectively. }
In all the systems the mean of the number of ring members was close to 6. Further, the variance in the  ring-size distribution was between $0.41$ and $0.49$ (see tab.\,\ref{table2}). 
{ This is a smaller range as reported by Tuan et al.\,\cite{Tuan2012} and  Kumar et al.\,\cite{Thorpe2012}, and below the values reported by Van Hoang\,\cite{VanHoang2015}}.
{ The top view of the structures of systems $\zeta_1,\ldots ,\zeta_6$ can be seen in fig.\,(\ref{topview}).}
The different sizes are indicated by colors. In all the systems the bonds of the 6-ring structures, which formed large connected structures, are colored gray. 5-atom rings are colored blue, 7-atom rings green and 4-, 8- and 9-atom rings red.

\begin{table}

\caption{Distribution of the coordination number of the first coordiation shell of the obtained systems $\zeta_1,\ldots ,\zeta_6$, and values reported using the WWW-method, MD-cooling and stochastic quench.{\color{black} The mean coordination number of the first coordination shell is given by $\langle n_1\rangle$.}}
\label{table3}
\begin{tabular}{|c|c|c|c|c|c|c|c|}
\hline
~ & 1 &  2 & 3 &4 &5 & 6 &  \color{black}$\langle n_1\rangle$ \\ \hline\hline
$\zeta_1$ & $0$ & $0.002$ & $0.988$ & $0.010$ & $0$ & $0$  & \color{black} $3.008$\\
$\zeta_2$ & $0$ & $0.002$ & $0.988$ & $0.010$ & $0$ & $0$ & \color{black}$3.008$  \\
$\zeta_3$ & $0$ & $0.002$ & $0.988$ & $0.010$ & $0$ & $0$ & \color{black}$3.008$ \\
$\zeta_4$ & $0$ & $0.010$ & $0.988$ & $0.002$ & $0$ & $0$ & \color{black}$2.992$\\
$\zeta_5$ & $0$ & $0.008$ & $0.984$ & $0.008$ & $0$ & $0$ & \color{black}$3.000$  \\
$\zeta_6$ & $0$ & $0.004$ & $0.988$ & $0.008$ & $0$ & $0$ & \color{black}$3.004$\\ \hline
{\bf WWW-meth.} &~&~&~&~&~&~&~\\
\begin{tabular}{@{}c@{}c@{}c@{}}Kapko\cite{Kapko2010}\\ Tuan\cite{Tuan2012}\\ Kumar\cite{Thorpe2012}\\ Popescu\cite{Popescu2013} \end{tabular} & $0$ & $0$ & $1.000$ & $0$ & $0$ & $0$ & $3$ \\
{\bf MD-cooling} &~&~&~&~&~&~&~\\
Kumar \cite{Thorpe2012} & $0$ & $0.006$ & $0.991$ & $0.003$  & $0$ & $0$ & $2.997$\footnote{value calculated from distribution of coordination number}\\
 Van Hoang \cite{VanHoang2015} & \begin{tabular}{@{}c@{}}$0.004$\\ $0.002$\end{tabular}&
\begin{tabular}{@{}c@{}} $0.083$ \\ $0.065$ \end{tabular} & 
\begin{tabular}{@{}c@{}} $0.457$ \\ $0.689$ \end{tabular} & 
\begin{tabular}{@{}c@{}} $0.356$ \\ $0.208$ \end{tabular} & 
\begin{tabular}{@{}c@{}} $0.091$ \\ $0.033$ \end{tabular} & 
\begin{tabular}{@{}c@{}} $0.009$ \\ $0.004$ \end{tabular} & 
\begin{tabular}{@{}c@{}} $3.474^{\it a}$ \\ $3.220^{\it a}$ \end{tabular} \\
{\bf stoch.\,quench} &~&~&~&~&~&~&~\\
Holmstr{\"o}m\cite{holmstrom2011} & $0.005$ & $0.010$ & $0.985$ & $0$ & $0$ & $0$ & $2.980^a$ \\ \hline
\end{tabular}
\end{table}

Fig.\,(\ref{radial}) shows a plot of the radial distribution function $g(r)$ of the systems $\{\zeta_1 ,\ldots ,\zeta_6\}$ (lines). The bars show the radial distribution function of a hexagonal graphene structure with a bond length of $1.45$\AA. The inset shows a magnification of the range from $4$ to $7$\AA . For the calculation of $g(r)$ the length of the intervals $\Delta r$ was set to $0.1$\AA \, for $0<r<3$\AA \, and $0.2$\AA \, for $r>3$\AA. 
With some small spread, the distribution of the radial distance between the atoms of the systems $\zeta_1,\ldots,\zeta_6$ clearly resembled the first four peaks (up to $4$\AA) in the radial distribution function of the graphene structure. From $4$ to $7$\AA \, only some of the systems showed, albeit relatively weak, the peaks of graphene. Which may indicate a weak medium range order. From $7$ to $10$\AA \, peaks were no longer distinctly visible indicative of the disordered nature of the system. Overall the six systems exhibited very similar radial distribution functions, { which are also qualitatively similar to the distributions reported in refs.\,\cite{Thorpe2012,holmstrom2011,Kapko2010,Li2011}.}

Fig.\,(\ref{bondangle}) shows the normalized distribution of bond angles of the different systems. The length of an interval was $\Delta \theta =1^o$. As can be seen, the most common angle was around $120.0^o$ corresponding to the internal angles of hexagon structures. A distinct second peak formed at $108.5^o$, corresponding to the pentagon structures [compare to fig.\,(\ref{rings})]. The peak for the heptagon angle (expected to be at $128.6^o$) was not distinct. Due to their larger size, 7-atom rings have more flexibility to bend, leading to a wider spread in the distribution of internal angles. This together with the close proximity to the hexagon angles explains why an expected peak of hexagon angles appears to be smoothed out.
The root mean square deviation of the bond angles of the systems from the mean can be seen in tab.\, (\ref{table1}). For the six systems $\{\zeta_1 ,\ldots,\zeta_6\}$ it was found to be in the range between $5.2^o$ and $5.8^o$. { These values are significantly smaller than the $\theta_{rms}$ values reported for the 2-dimensional structures in refs.\,\cite{Kapko2010,Tuan2012,Thorpe2012}.  Peaks at $60^o$ and $90^o$, as reported by Holmstr{\"o}m et al.\,\cite{holmstrom2011} were not observed. However, for his 3-dimensional structure Holmstr{\"o}m also noted the highest probability to be around the $120^o$ angle, similar to the present study [see fig.(\ref{bondangle})]. }\\
The coordination numbers of the first coordination shell (counting bonds up to a radius of $2.0$\AA\,) were very close to 3.
{ They were dominated by 3 bonds per atom [see tab.\,(\ref{table2})], which can be explained by the preference of the carbon atoms in single layer sheets to bond to three neighbors. The percentage of atoms to form 2 or 4 bonds was very low, i.e. combined $\leq 1.6\%$. This is in excellent agreement with refs.\,\cite{Kapko2010,Tuan2012,Thorpe2012,Popescu2013}. It is to note that the WWW-method is essentialy a bond-rotation algorithm and hence confines the number of bonds per atom to exactly 3. Only the systems by Van Hoang\,\cite{VanHoang2015} and Holmstr\"om\,\cite{holmstrom2011} showed single coordinated atoms, and Van Hoangs\,\cite{VanHoang2015} distributions were generally wider spread.} \\
The average energies per atom $U/N_p$ of the systems were very similar, with a mean of $-7.073 eV$ and a standard deviation of $\sigma_U=0.013 eV$. See tab.\,(\ref{table1}), $\zeta_4$ had the lowest energy of $-7.089 eV$ and $\zeta_2$ the highest of $-7.050 eV$. \\
\begin{table}[t!]
\caption{Table of the average energy per atom $U/N_p$, the height $\Delta z$ and root mean square height $rms(z)$.}
\label{table1}
\begin{tabular}{|c|c|c|c|c|c|c|}
\hline
~ & $\zeta_1$ &  $\zeta_2$ & $\zeta_3$ & $\zeta_4$ & $\zeta_5$ & $\zeta_6$    \\ \hline\hline
 $U/N_p$ (eV) & $-7.067$ & $-7.050$ & $-7.086$ & $-7.089$ & $-7.070$ & $-7.073$ \\
$\Delta z$ (\AA) & $3.75$ & $4.64$ & $4.16$ & $5.30$ & $4.40$ & $4.07$ \\
$rms(z)$ (\AA) & $0.81$ & $0.90$ & $0.79$ & $1.14$ & $0.78$ & $0.84$\\
\hline
\end{tabular}
\end{table}
The physical width of the obtained structures was calculated with respect to a plane determined by the location of all the atoms. The smallest width was found for $\zeta_1$ as $\Delta z=3.75$\AA\, and the largest for $\zeta_4$ as $\Delta z=5.30$\AA . The root mean square value of the width  was found to be between $0.78$\AA\, ($\zeta_5$) and $1.14$\AA\, ($\zeta_4$). 
Puckering in single layer amorphous graphene sheets has also been observed in refs.~\cite{Li2011,Popescu2013,Mortazavi2016,LiandDrabold2013,holmstrom2011}.
The systems in refs.\,\cite{Kapko2010,Tuan2012,Thorpe2012} were confined to two dimensions, which prevented the structures to pucker.
 \\
Overall, while representing different structural arrangements of the atoms, the systems obtained with the described optimization procedure showed very similar properties.

\section{Conclusion}
A temperature driven differential mutation method was introduced.
The method combines a differential mutation algorithm with a thermal selection criteria
for real-space optimization of amorphous systems using an anlytical expression for the energy stored in the atomic configurations.
 The inclusion of the temperature in the selection criteria allows for the breaking of existing bonds in order to rearrange the atoms and form new bonds. This corresponds to a climb over energy barriers, where the systems temporarily are allowed to take energetically less favorable configurations. This in turn allows the systems to escape local minima.
 The method is capable to deal with structural optimization of amorphous materials using only a small population size consisting of the different systems. It reliably obtains low energy structures corresponding to very low minima on the energy landscape.\\ 
The method was tested on the structural optimization of amorphous graphene, starting from unbiased randomly selected locations of the atoms. The population consisted of 6 systems $\{\zeta_1,\ldots ,\zeta_6\}$, each with $500$ atoms.
The results showed that, while being microscopically distinct different structures, the overall properties of the obtained systems were very similar. No distinct differences between $\zeta_1,\zeta_2,\ldots$ and $\zeta_6$ could be observed in the radial distribution function, the distribution of bond angles, average energy per atoms and coordination number. Some minor difference in the distribution of rings could be observed. However, it is to be expected that if the size of the systems were to increase these differences might also vanish.

\end{document}